\documentclass[12pt]{article}

\usepackage{amssymb}

\newcommand{\be}{\begin{equation}}
\newcommand{\ee}{\end{equation}}
\newcommand{\bea}{\begin{eqnarray}}
\newcommand{\eea}{\end{eqnarray}}
\def\tr{{\rm tr}}

\def\dint{\displaystyle \int }

\def\cN{{\cal N}}
\newcommand{\m}{d\zeta(^{33}_{11})du\,}

\begin{document}
 \begin{center}
 {\Large\bf
On Low-Energy Effective Action in $\cN=3$ Supersymmetric Gauge Theory
}
 \\[0.5cm]
 {\bf
 I.B. Samsonov\footnote{email: samsonov@phys.tsu.ru}\\
\it Laboratory of Mathematical Physics, Tomsk Polytechnic
 University,\\ 30 Lenin Ave, Tomsk 634034, Russia
 }
 \\[0.8cm]
\bf Abstract
\end{center}
\begin{quotation}\small\noindent
The problem of construction of low-energy effective action in $\cN=3$ SYM
theory is considered within the harmonic superspace (HSS) approach. The low-energy
effective action is supposed to be a gauge- and scale-invariant functional
in $\cN=3$ HSS reproducing the term $F^4/\phi^4$ in components. This
functional is found as a scale-invariant generalization of the $F^4$-term
in $\cN=3$ supersymmetric Born-Infeld action.
\end{quotation}

\setcounter{footnote}{0}

 \section{Introduction}
Extended supersymmetric field theories play the important role in modern
high-energy theoretical physics due to their beautiful
classical and quantum properties and close relations to string/brane
theory.  One of the most popular examples is $\cN=4$ gauge theory,
quantum aspects of which attracts much attention.  The symmetries of
this theory are so rich ($\cN=4$ superconformal symmetry) that many
properties of this model can be proved on the symmetry grounds only.

In this paper we study the model of $\cN=3$ super Yang-Mills formulated in
$\cN=3$ harmonic superspace \cite{Harm1}. This model, like $\cN=4$ SYM, is
known to be finite \cite{Delduc} and superconformally invariant \cite{Harm2}.
Moreover, $\cN=3$ SYM model describes the dynamics of the same multiplet of
physical field as $\cN=4$ one (see e.g. book
\cite{book2} for a review). Therefore it can be
considered as an alternative off-shell extension of $\cN=4$ model while
on-shell they both have equivalent dynamics. However, the structure of low-energy
effective action in $\cN=3$ SYM model was not studied as yet.
As another evidence in favour of usefulness of the $\cN=3$ HSS
approach, it was recently shown \cite{Grassi} that $\cN=3$ SYM theory
in harmonic superspace is naturally generated from superstring theory.

It is well known \cite{Dine,BuchN4} that the leading term in the
low-energy effective action of $\cN=4$ SYM model in the sector of $\cN=2$
vector multiplet has the form $\int d^4x\frac{F^2\bar
F^2}{(\bar\phi\phi)^2}$, where
$F_{\alpha\beta}=\partial_{\alpha\dot\alpha}A^{\dot\alpha}_\beta+
\partial_{\beta\dot\alpha}A^{\dot\alpha}_\alpha$ is the Abelian strength field and $\phi$ is
the scalar field corresponding to $\cN=2$ vector multiplet. Complete
$\cN=4$ supersymmetric generalization of the low-energy effective action
including both vector fields and hypermultiplets was given in
\cite{BuchIvanov}, the leading bosonic component of this action is
\be
\int d^4x \frac{F^2\bar F^2}{(\bar \phi_i\phi^i)^2},
\label{F4}
\ee
where $\phi^i$ is SU(3) triplet of scalar fields. Since the
models of $\cN=3$ and $\cN=4$ SYM are equivalent on-shell, we expect that
the term (\ref{F4}) is also leading in the effective action of $\cN=3$ SYM
model. Therefore it is interesting to find $\cN=3$ superfield action
reproducing the expression (\ref{F4}) in the component expansion. One can
expect that such an action corresponds to the low-energy effective
action of $\cN=3$ SYM model.

An important step in understanding the possible structure of the
effective action in $\cN=3$ gauge theory was the construction of $\cN=3$
supersymmetric Born-Infeld action \cite{Zupnik01} where it was shown that
there exists a natural description of the $F^4$  term (and all higher-order ones) by
unconstrained $\cN=3$ superfields in harmonic superspace.
 We suppose that the low-energy effective
action of $\cN=3$ gauge theory should correspond to a scale invariant
generalization of the corresponding BI-theory. In this work we propose some
possible form of such a functional satisfying the conditions of
supersymmetric, gauge and scale invariances and reproducing the term
(\ref{F4}) in components.

This talk is a review of our recent results published in
\cite{Samsonov04}.

\section{$\cN=3$ multiplets and actions in harmonic superspace}
The $\cN=3$ HSS was introduced in ref. \cite{Harm1} to construct
an off-shell superfield formulation of $\cN=3$ SYM model. The basics
of the harmonic superspace method are exposed in book \cite{IvanovBook}.
Throughout this paper we follow the conventions of recent works
\cite{Zupnik01,Samsonov04}.

The $\cN=3$ HSS \cite{Harm2,IvanovBook} is defined as the superspace with
coordinates $\{Z,u\}$, where
$Z=\{x^{\alpha\dot\alpha},\theta_i^\alpha,\bar\theta^{i\dot\alpha} \}$
\footnote{We denote by small Greek symbols the SL(2,$C$)
spinor indices, $\alpha,\dot\alpha,\ldots=1,2$;
the small Latin letters are $SU(3)$ indices, $i,j,\ldots=1,2,3$.}
is a set of standard $\cN=3$ coordinates and $u$ are the harmonics parameterizing
the coset $SU(3)/U(1)\times U(1)$. We consider the harmonics $u^I_i$ and their
conjugate $\bar u_I^i$ ($I=1,2,3$) as $SU(3)$ matrices
\be
u^I_i\bar u_J^i=\delta^I_J,\qquad
u^I_i\bar u_I^j=\delta_i^j,\qquad
\varepsilon^{ijk}u^1_iu^2_ju^3_k=1.
\label{e1}
\ee

The harmonic superspace $\{Z,u\}$ contains the so called analytic subspace with
the coordinates $\{\zeta_A,u\}=\{x_A^{\alpha\dot\alpha},\theta_2^\alpha,
\theta_3^\alpha,\bar\theta^{1\dot\alpha},\bar\theta^{2\dot\alpha},u\}$ where
\be
x_A^{\alpha\dot\alpha}=x^{\alpha\dot\alpha}-
 2i(\theta_1^\alpha\bar\theta^{1\dot\alpha}-
 \theta_3^\alpha\bar\theta^{3\dot\alpha})\,, \qquad
\theta_I^\alpha=\theta_i^\alpha \bar u^i_I\,, \qquad
\bar\theta^{I\dot\alpha}=\bar\theta^{\dot\alpha i}u^I_i\,.
\label{e2}
\ee
The analytic superspace plays an important role in harmonic superspace
approach since it is closed under supersymmetry and $\cN=3$ SYM action is
written in analytic coordinates.

The harmonic superspace is equipped with Grassmann covariant derivatives
$D^I_\alpha,\bar D_{I\dot\alpha}$ and harmonic covariant ones $D^I_J$ which
form the $su(3)$ algebra (see \cite{Harm2,IvanovBook} for details). For
example, the derivatives which define the analytic superfields are
\be
D^1_\alpha=\frac\partial{\partial\theta_1^\alpha},\qquad
\bar D_{3\dot\alpha}=-\frac\partial{\partial\bar\theta^{3\dot\alpha}}.
\label{D_}
\ee
Acting on any analytic superfield, the derivatives (\ref{D_}) give zero.
This is similar to the chiral superfields which are annihilated by
the corresponding chiral derivatives.

There are two multiplets of analytic superfields which are important for
us. The first one is the multiplet of $\cN=3$ gauge prepotentials $V^1_2$,
$V^2_3$, which are complex (mutually conjugate) analytic superfields.
Physical bosonic component fields are contained in the prepotentials as
\cite{Harm1}
\be
\begin{array}{rl}
V^2_3=&[(\bar\theta^1\bar\theta^2)u^2_k-(\bar\theta^2)^2u^1_k]\phi^k
+\theta_3^\alpha\bar\theta^{2\dot\alpha}A_{\alpha\dot\alpha}
-i\theta_2^\alpha\theta_3^\beta(\bar\theta^2)^2H_{\alpha\beta}\\
&+{\rm spinors\ and\ auxiliary\ fields},\\
V^1_2=&-\widetilde{(V^2_3)}=-[(\theta_2\theta_3)\bar u_2^k-(\theta_2)^2\bar
u_3^k]\bar\phi_k+\theta_2^\alpha\bar\theta^{1\dot\alpha}A_{\alpha\dot\alpha}
+i(\theta_2)^2\bar\theta^{1\dot\alpha}\bar\theta^{2\dot\beta}
 \bar H_{\dot\alpha\dot\beta}\\
&+{\rm spinors\ and\ auxiliary\ fields}.
\end{array}
\label{e4}
\ee
Here, $\phi^i,\bar\phi_i$ are complex scalar fields,
$A_{\alpha\dot\alpha}$ is a vector gauge field, $H_{\alpha\beta}$,
$\bar H_{\dot\alpha\dot\beta}$ are the auxiliary fields which ensure
the correct structure of the gauge field sector of the theory \cite{Zupnik01}.
The prepotentials $V^1_2$, $V^2_3$ are used in the formulation of $\cN=3$
SYM model in harmonic superspace. For example, the quadratic (free)
$\cN=3$ SYM action is
\be
S_2[V]=-\frac14\tr\dint\m[V^2_3D^1_3V^1_2+\frac12(D^1_2V^2_3-D^2_3V^1_2)^2].
\label{S2_}
\ee
The integration in (\ref{S2_}) is performed over analytic superspace
($\m$ is the integration measure on $\cN=3$ analytic HSS).

The component form of the action $S_2$ in the sector of gauge fields
is \cite{Zupnik01}
\be
S_2=\int d^4x[V^2+\bar V^2-2(\bar V\bar F+ VF)+\frac12(F^2+\bar F^2)],
\label{S2}
\ee
where
\be
\begin{array}l
V_{\alpha\beta}=\frac14(H_{\alpha\beta}+F_{\alpha\beta}),\qquad
\bar V_{\dot\alpha\dot\beta}=\frac14(\bar H_{\dot\alpha\dot\beta}+
 \bar F_{\dot\alpha\dot\beta}),\\
F^2=F^{\alpha\beta}F_{\alpha\beta},\quad
V^2=V^{\alpha\beta}V_{\alpha\beta},\quad
FV=F^{\alpha\beta}V_{\alpha\beta}.
\end{array}
\ee
The auxiliary fields $V_{\alpha\beta}$, $\bar V_{\dot\alpha\dot\beta}$ can be
eliminated by their algebraic classical equations of motion
\be
V_{\alpha\beta}=F_{\alpha\beta},\qquad
\bar V_{\dot\alpha\dot\beta}=\bar F_{\dot\alpha\dot\beta}.
\label{aux}
\ee
As a result, the free classical action (\ref{S2}) takes the form of the usual
Maxwell action
\be
S_2=-\frac12\int d^4x(F^2+\bar F^2).
\label{maxwell}
\ee

Another important $\cN=3$ multiplet is described by $\cN=3$ superfield
strengths which are expressed through prepotentials as
\be
\begin{array}{ll}
W_{23}=\frac14\bar D_{3\dot\alpha}\bar D_3^{\dot\alpha}V^3_2\,,\qquad&
\bar W^{12}=-\frac14D^{1\alpha}D^1_\alpha V^2_1,\\
W_{12}=D^3_1 W_{23},& \bar W^{23}=-D^3_1\bar W^{12},\\
W_{13}=-D^2_1W_{23},& \bar W^{13}=D^3_2 \bar W^{12}.
\end{array}
\label{str}
\ee
Here $V^2_1$, $V^3_2$ are non-analytic prepotentials which are the
solutions of zero-curvature equations \cite{Zupnik01}
\be
D^2_1 V^1_2= D^1_2 V^2_1,\qquad
D^3_2 V^2_3= D^2_3 V^3_2.
\label{zero-curv}
\ee
The superfields (\ref{str}) have the following component structure
in the sector of physical bosons
\cite{Ferrara}
\be
\begin{array}{rl}
W_{23}=&u^1_i\phi^i(x_{A+})+4i\theta_2^\alpha\theta_3^\beta
V_{\alpha\beta}(x_{A+})
+{\rm spinors\ and\ auxiliary\ fields},\\
\bar W^{12}=&\bar u_3^i\bar\phi_i(x_{A-})
+4i\bar\theta^{1\dot\alpha}\bar\theta^{2\dot\beta}
\bar V_{\dot\alpha\dot\beta}(x_{A-})
+{\rm spinors\ and\ auxiliary\ fields},
\end{array}
\label{e10}
\ee
where $x_{A\pm}^{\alpha\dot\alpha}=x_A^{\alpha\dot\alpha}\pm
2i\theta_2^\alpha\bar\theta^{2\dot\alpha}$.

The strength superfields (\ref{str}) are used in construction of
$\cN=3$ supersymmetric Born-Infeld action \cite{Zupnik01}.
For example, the quartic term of the $\cN=3$ BI action is described by the
following superfield action
\be
S_4=\frac1{32}\dint\m \frac{(\bar W^{12}W_{23})^2}{X^2}.
\label{S4_}
\ee
This action produces the first nontrivial term
\be
\frac12\int d^4x \frac{F^2\bar F^2}{X^2}
\label{F44}
\ee
in the Born-Infeld action.

\section{Construction of the leading term in $\cN=3$ SYM
low-energy effective action}
In this Section we construct a manifestly $\cN=3$ supersymmetric low-energy
effective action containing the term $F^4/{\phi}^4$ in the bosonic sector.

$\cN=3$ SYM theory is known to be a superconformal field theory
\cite{Harm2}, like the $\cN=4$ SYM one. Moreover, both these models
describe the dynamics of the same multiplet of physical fields and
therefore are on-shell equivalent \cite{book2}. The
effective action of $\cN=3$ SYM model should be scale invariant.
The transformations of dilatations (scale invariance) and $\gamma_5$-symmetry
(R-symmetry) act on the coordinates of harmonic superspace and superfield strengths as follows
\bea
&&\delta x^m_A=ax^m_A,\quad\delta\theta^\alpha_I=\frac12(a+ib)\theta^\alpha_I,
\quad\delta\bar\theta^{I\dot\alpha}=\frac12(a-ib)\bar\theta^{I\dot\alpha}
\nonumber\\
&&\delta W_{IJ}=(-a+ib)W_{IJ},\quad \delta\bar W^{IJ}=-(a+ib)\bar W^{IJ}.
\eea
We expect that a scale and $\gamma_5$-invariant generalization of the action
(\ref{S4_}) should correspond to the low-energy effective action of $\cN=3$
SYM model. In components such an action should reproduce
the scale and $\gamma_5$-invariant generalization of (\ref{F44}), that is (\ref{F4}).
Note that exactly this term is leading in the low-energy effective action
of $\cN=4$ SYM model \cite{Dine,BuchN4}.
Thus we wish to construct a generalization of the action
(\ref{S4_}) which would respect the scale- and $\gamma_5$-invariances.

To pass from (\ref{F44}) to the scale invariant component action (\ref{F4}),
one should replace the dimensionful constant $X$ by
the function of scalar fields $(\phi^i\bar\phi_i)$. Therefore, to obtain
a scale invariant generalization of the superfield action (\ref{S4_}) we have to replace the
constant $X$ by some superfield expression having the
same dimension and containing $\phi^i\bar\phi_i$ as the lowest component.
The suitable expression is
\be
\bar W^{IJ}W_{IJ}=\bar W^{12}W_{12}+\bar W^{23}W_{23}+\bar W^{13}W_{13}\,.
\label{WW}
\ee
Indeed, the component expansion of this superfield starts with the scalars
(see \cite{Ferrara} for details)
\be
\bar W^{IJ}W_{IJ}|_{\theta=\bar\theta=0}=\phi^i\bar\phi_i\,.
\label{WW|}
\ee

The expression (\ref{WW}) cannot be naively put into the
integral in (\ref{S4_}) in place of the constant $X$. The reason is that
the superfield $(\bar W^{IJ}W_{IJ})$ is not analytic since the superfield strengths
$\bar W^{23},\bar W^{13},W_{12},W_{13}$ are not analytic, while the
integration in (\ref{S4_}) goes over the analytic superspace. Therefore we
have to rewrite the action (\ref{S4_}) in full $\cN=3$ HSS and then to insert
$\bar W^{IJ}W_{IJ}$ into the integral.

The action (\ref{S4_}) in the full $\cN=3$ HSS is written as
\be
S_4=\frac{1}{32}\int d^4x d^{12}\theta du \frac1{X^2}\left[\frac{(\bar
D_1)^2}{4\square}(W_{23})^2\right]
\left[\frac{(D^3)^2}{4\square}(\bar W^{12})^2\right].
\label{S4full}
\ee
Replacing the constant $X$ by the superfield $\bar W^{IJ}W_{IJ}$
in (\ref{S4full}), we arrive at the action
\be
S_{4}^{scale-inv}=\alpha\int d^4xd^{12}\theta du\frac 1{(\bar W^{IJ}W_{IJ})^2}
\left[\frac{(\bar D_1)^2}{4\square}(W_{23})^2\right]
\left[\frac{(D^3)^2}{4\square}(\bar W^{12})^2\right],
\label{Seff}
\ee
where $\alpha$ is some {\it dimensionless} constant.
This constant cannot be fixed on the symmetry grounds only.
One of the possible ways of finding $\alpha$ is
a straightforward calculation of low-energy
effective action in the framework of quantum field theory.
Since the action (\ref{Seff}) includes no any dimensional constants, it is
scale invariant.

Let us study the component structure of the action (\ref{Seff}).
Note that the superfield strengths entering the action contain a multiplet
of physical fields as well as an infinite number of auxiliary fields.
We are interested in the component structure of the action (\ref{Seff}) in
the sector of scalar and vector physical fields. For this purpose
we neglect all the derivatives of scalar fields and Maxwell field strength.
Such an approximation is sufficient for retrieving the term $F^4/\phi^4$
while going to components. Therefore we use the following ansatz for the superfield strengths
\be
\begin{array}{ll}
\hat{\bar W}{}^{12}=\bar\phi_3+\bar\omega^{12}, & \hat W_{12}=\phi^3+\omega_{12},\\
\hat{\bar W}{}^{23}=\bar\phi_1+\bar\omega^{23}, & \hat W_{23}=\phi^1+\omega_{23},\\
\hat{\bar W}{}^{13}=-\bar\phi_2+\bar\omega^{13}, & \hat
W_{13}=-\phi^2+\omega_{13},
\end{array}
\label{str-comp}
\ee
where
\be
\begin{array}{ll}
\bar\phi_I=\bar u_I^i\bar\phi_i,& \phi^I=u^I_i\phi^i\,,\\
\bar\omega^{IJ}=4i\bar\theta^{I\dot\alpha}\bar\theta^{J\dot\beta}
 \bar V_{\dot\alpha\dot\beta}\,,&
\omega_{IJ}=4i\theta_I^\alpha\theta_J^\beta V_{\alpha\beta}.
\end{array}
\label{str-comp1}
\ee
The symbol ``hat''
indicates that we consider only scalar and vector bosonic fields and discard any
auxiliary fields.
In the ansatz (\ref{str-comp}) and (\ref{str-comp1}) the action
(\ref{Seff}) contains only local terms
\be
\hat S_{4}^{scale-inv}=\alpha\int d^4x d^{12}\theta du \frac{(\theta_1)^2(\bar\theta^3)^2}{
(\hat{\bar W}{}^{IJ}\hat W_{IJ})^2}(\hat{\bar W}{}^{12}\hat W_{23})^2.
\label{S1}
\ee
Performing the integration over Grassmann and harmonic variables in
(\ref{S1}), we obtain
\be
\hat S_4^{scale-inv}=\frac{\alpha_0}2\int d^4x\frac{V^2\bar
V^2}{(\phi^i\bar\phi_i)^2},
\label{S4}
\ee
where $\alpha_0=\frac{32}{15}\alpha$.

Now we should express the auxiliary fields $V_{\alpha\beta},
\bar V_{\dot\alpha\dot\beta}$ through the physical field strengths
$F_{\alpha\beta}, \bar F_{\dot\alpha\dot\beta}$ from the action
\be
\hat S_2+\hat S_{4}^{scale-inv}=\int d^4x\left[V^2+\bar V^2-2(\bar V\bar F+ VF)+
\frac12(F^2+\bar F^2)+
\frac{\alpha_0}2\frac{V^2\bar V^2}{(\phi^i\bar\phi_i)^2}\right].
\label{S5}
\ee
The corresponding equations of motion for the auxiliary fields
$V_{\alpha\beta}$, $\bar V_{\dot\alpha\dot\beta}$ are
\be
2F_{\alpha\beta}=V_{\alpha\beta}\left[2+\frac{\alpha_0}{(\phi^i\bar\phi_i)^2}\bar
V^2\right],\qquad
2\bar F_{\dot\alpha\dot\beta}=\bar V_{\dot\alpha\dot\beta}
\left[2+\frac{\alpha_0}{(\phi^i\bar\phi_i)^2}V^2\right].
\label{EOM}
\ee
Eqs. (\ref{EOM}) define the auxiliary fields $V_{\alpha\beta},
\bar V_{\dot\alpha\dot\beta}$ as functions of
$F_{\alpha\beta}, \bar F_{\dot\alpha\dot\beta}$. The solution to these
equations can be represented as a series over the Maxwell field strengths:
\be
V_{\alpha\beta}=F_{\alpha\beta}\left[1-\frac{\alpha_0}{2(\phi^i\bar\phi_i)^2}
 \bar F^2+O(F^3)\right],\quad
\bar V_{\dot\alpha\dot\beta}=\bar F_{\dot\alpha\dot\beta}\left[1-
\frac{\alpha_0}{2(\phi^i\bar\phi_i)^2}F^2+O(F^3)\right].
\label{solEOM}
\ee
Substituting the solutions (\ref{solEOM}) into the action (\ref{S4}),
we find
\be
S_{4}^{scale-inv}=\frac{1}2\int d^4x \left[\frac{F^2\bar F^2}{(\phi^i\bar\phi_i)^2}
-\frac{1}2\frac{F^2\bar F^2}{(\phi^i\bar\phi_i)^4}(F^2+\bar F^2)
+O(F^8)
\right],
\label{S6}
\ee
where we set $\alpha_0=1$ for simplicity. As a result, we see that the
main bosonic component in the action (\ref{Seff}) is exactly the term
$F^4/\phi^4$ (\ref{F4}) which is the first nontrivial term in the $\cN=3$ SYM
model. The action (\ref{S6}) contains also all higher-order terms starting with $F^6$.
However, the consideration of these terms requires the special attention (the
corresponding analysis is performed in \cite{Samsonov04}).

Let us finish this Section with several comments concerning the
superfield action (\ref{Seff}).

\begin{itemize}
\item This action contains the nonlocal operator
$\square^{-1}$.  However, as is shown above, the leading low-energy term in the component
action is local.
\item
 From the very beginning there is a freedom in distributing the derivatives
 among different factors in the actions (\ref{S4full}) and (\ref{Seff}).
 However, the local part of the action (\ref{Seff}) actually does not depend
 on the specific pattern of such a distribution.
\item As follows from eq. (\ref{S6}), the action $S_4^{scale-inv}$
contains the term (\ref{F4}) in its component
expansion. We observe that in this expression the scalar fields appear in a single
$SU(3)$ invariant combination. An analogous result was earlier obtained in
ref. \cite{BuchIvanov} for the full low-energy effective action of
$\cN=4$ SYM in the $\cN=2$ HSS approach.
The advantage of $\cN=3$ formalism is that all scalar fields
from the very beginning are included into a single
$\cN=3$ multiplet, while in the $\cN=2$ superspace language the scalar
fields are distributed between vector multiplet and hypermultiplet.
\item
The off-shell action (\ref{Seff}) is manifestly supersymmetric, gauge
invariant and scale invariant. It also respects the invariance under the
$\gamma_5$ and $SU(3)$ transformations. Therefore, it can be considered as
a candidate for the low-energy effective action in $\cN=3$ SYM model.
\end{itemize}

 \section{Summary}
 In this paper we analyzed the possible off-shell structure of
 low-energy effective action of $\cN=3$ SYM model written in
 $\cN=3$ harmonic superspace. This action was obtained as an $\cN=3$ superfield
 generalization of the term $F^4/\phi^4$ which is leading in the low-energy
 effective action. It is written as a functional built out of
 the superfield strengths in full $\cN=3$ superspace. This functional is
 manifestly supersymmetric, gauge invariant, scale and $\gamma_5$-invariant and
 corresponds to a scale invariant generalization of 4-th order term in
 the $\cN=3$ supersymmetric BI action.

In conclusion, let us point out once again that the effective action (\ref{Seff}) was
found solely by employing the symmetries of the model and the requirement
that it produces the $F^4/\phi^4$ term in components.
This action was determined up to an arbitrary numerical coefficient.
The important problem now is to reproduce the action (\ref{Seff})
by direct quantum field theory computations in $\cN=3$ HSS.

\subsubsection*{Acknowledgements}
I am very grateful to the organizers of the conference QUARKS-2004 for the
invitation and partial financial support of my participation.
I should like to thank Prof. I.L. Buchbinder for many constructive discussions
during this work and guiding advices which helped me to prepare
this talk properly. I am grateful also to Prof. E.A. Ivanov and
Prof. B.M. Zupnik which explained me many important details of the harmonic
superspace approach and took part in this research.
The work was partially supported by RFBR grant, project No 03-02-16193;
INTAS grant, project No 00-00254; LSS grant, project No 1743.2003.2.

 \end{document}